\documentclass{svjour3}                     
\usepackage{graphicx}
\usepackage{mathptmx}      

\usepackage{amsmath}
\usepackage{amsfonts}
\usepackage{amssymb}

\newcommand{\sqrts}{\sqrt{s}}
\newcommand{\sqrtsnn}{\sqrt{s_{_{\ensuremath{\it{NN}}}}}}
\newcommand{\sqrtsgzk}{\sqrt{s_{_{\ensuremath{\it{GZK}}}}}}
\newcommand{\meanptgzk}{\ensuremath{\left< p_{\perp} \right>}_{_{\ensuremath{\it{GZK}}}}}

\newcommand{\pp}{$p$-$p$}
\newcommand{\ppbar}{$p$-$\bar p$}

\newcommand{\Lambdaqcd}{\Lambda_{\ensuremath{\it QCD}}}

\newcommand{\epos}{\textsc{epos}}
\newcommand{\qgsjet}{\textsc{qgsjet}} 
\newcommand{\sibyll}{\textsc{sibyll}}
\newcommand{\pythia}{\textsc{pythia}}

\newcommand{\dNdeta}{dN_{ch}/d\eta|_{\eta=0}}
\newcommand{\meanpt}{\ensuremath{\left< p_{\perp} \right>}}
\newcommand{\pT}{p_{\perp}}

\journalname{Few Body Systems}
\begin{document}

\title{The strong interaction at the collider and cosmic-rays 
frontiers\thanks{Presented at the workshop "30 years of strong interactions", Spa, Belgium, 6-8 April 2011.}}

\titlerunning{The strong interaction at collider and cosmic-rays energies}        

\author{David d'Enterria \and
Ralph Engel \and Tanguy Pierog \and
Sergey Ostapchenko \and Klaus Werner
}


\institute{
D. d'Enterria \at CERN, PH Department, 1211 Geneva, Switzerland \\
              \email{dde@cern.ch}           
\and
R. Engel, T. Pierog \at Karlsruhe Institut of Technology, Postfach 3640, 76021 Karlsruhe, Germany
\and
S. Ostapchenko \at NTNU, Inst. for Fysikk, 7491 Trondheim, Norway
\and
K. Werner \at SUBATECH, 4 rue Alfred Kastler, BP 20722, 44307 Nantes Cedex 3, France
}

\date{Received: date / Accepted: date}

\maketitle

\begin{abstract}
  First data on inclusive particle production measured in proton-proton collisions 
  at the Large Hadron Collider (LHC) 
  are compared to predictions of various hadron-interaction Monte Carlos 
  (\qgsjet, \epos\ and \sibyll) used commonly in high-energy cosmic-ray physics. 
  While reasonable overall agreement is found for some of the models, 
  none of them reproduces consistently the $\sqrts$ evolution of all the
  measured observables. We discuss the implications of the new LHC data for the
  modeling of the non-perturbative and semihard parton dynamics in hadron-hadron 
  and cosmic-rays interactions at the highest energies studied today.

\end{abstract}

\section{Introduction}
\label{sec:intro}

The highest energy hadronic interactions measured on Earth result from 
the collision of cosmic rays (CR) -- protons and nuclei accelerated in various 
astrophysical sources that propagate through the universe up to the so-called 
Greisen-Zatsepin-Kuzmin (GZK) cutoff around $10^{20}$~eV~\cite{Greisen:1966jv,Zatsepin66e} -- 
with air nuclei in the upper atmosphere~\cite{Bluemer:2009zf}.
The determination of the primary energy and identity (mass) of such ultra-high-energy 
cosmic rays relies on the study of the cascade of secondary particles,
called extensive air-showers (EAS), that they produce in the atmosphere~\cite{Knapp:2002vs},
and its comparison to simulations 
that include the modeling of hadronic 
interactions at c.m. energies ($\sqrtsgzk\approx 400$\,TeV) 
more than two orders of magnitude higher than those studied at particle colliders
before the LHC. The dominant source of uncertainty in the interpretation of the highest-energy 
EAS data stems from our limitations to model particle production in 
strongly interacting systems. Indeed, even at asymptotically large energies the 
collision between two hadronic objects is sensitive to non-perturbative 
-- hadronization, beam remnants, soft ``peripheral'' diffractive scatterings --
or semi-hard -- saturation of gluon densities, multi-parton interactions --
dynamics that need still to be constrained directly from experimental data.\\

The first LHC data have extended by more than a factor of 
three the c.m.~energies for which we have direct proton-proton measurements available 
to test and constrain the ingredients of the hadronic Monte Carlo (MC) codes used in CR 
physics. In this work we compare the predictions of several CR MCs 
with various inclusive observables measured at the LHC which are
sensitive to non-perturbative and semihard QCD dynamics:
\begin{description}
\item (i) total inelastic \pp\ cross sections $\sigma_{inel}$, 
\item (ii) pseudorapidity density of charged particles at midrapidity $\dNdeta$, 
\item (iii) event-by-event distribution of the charged particle multiplicity $P(N_{ch})$,
\item (iv) energy distribution of (very) forward particles $dN/dE_{\gamma }|_{|\eta|>10.94}$ and $dE_{had}/d\eta|_{|\eta|=3-5}$,
\item (v) average transverse momentum of the produced hadrons $\meanpt$. 
\end{description}
The implications of the data-theory comparisons for the improvement of the description of 
multiparticle production in the hadronic event generators and for the interpretation of CR results are discussed.
The interested reader can find more details in~\cite{DdE:2011,Pierog2011}.

\section{Hadronic collisions at multi-TeV energies}
\label{sec:hadronMCs}

The inclusive production of particles in high-energy hadronic
collisions receives contributions from ``soft'' and ``hard''
interactions between the partonic constituents of the colliding hadrons. 
Soft (resp. hard) processes involve mainly $t$-channel partons of 
virtualities $q^2$ typically below (resp. above) a scale $Q_0^2$ of 
a few~GeV$^2$. 

\paragraph{Soft scatterings} give rise to production of hadrons with small 
transverse momenta $\pT$ and dominate hadronic collisions at low energies 
($\sqrts\lesssim$~20~GeV). Although soft processes have a virtuality scale 
not far from $\Lambdaqcd \approx$~0.2~GeV and thus cannot be treated within 
perturbative QCD (pQCD), predictions based on basic quantum field-theory
principles -- such as unitarity and analyticity of scattering amplitudes --
as implemented in the Gribov's Reggeon Field Theory (RFT)~\cite{Gribov:1968fc}, 
give a decent account of their cross sections in terms of the exchange of
virtual quasi-particle states (Pomerons and Reggeons). At high energies
the dominant soft contributions are from diffractive scatterings where
one or both colliding hadrons survive the interaction and few particles
are produced.

\paragraph{(Semi)hard parton-parton scatterings} dominate the inelastic hadron production 
cross-sections for c.m.~energies above a few hundreds of GeV. Hard processes 
with large $|q^2|\gg \Lambdaqcd^2$ can be treated within perturbative QCD in a 
collinear-factorized approach in terms of parton distribution functions (PDFs)
in the hadron convoluted with the elementary parton-parton subprocess computable
at a given order in the 
strong coupling constant $\alpha_s(q^2)$. 
The scattered quarks and gluons produce then collimated bunches of final-state 
hadrons (jets) in a branching process dominated by perturbative parton splittings 
described by the Dokshitzer-Gribov-Lipatov-Altarelli-Parisi (DGLAP) 
equations~\cite{Gribov:1972ri,Altarelli77,Dokshitzer77}, and by non-perturbative 
hadronization (e.g. based on the Lund string model~\cite{Andersson:1983ia})
when the parton virtuality is below $\cal{O}$(1~GeV).
At increasingly larger c.m.~energies, one needs to account for multi-parton 
scatterings and parton saturation effects. First, the cross section predicted 
by the (semi)hard processes exceeds the total inelastic \pp\ cross section
for $\pT$ values of a few GeV indicating that multiple parton interactions (MPI) 
occur per collision. Second, for decreasing but still perturbative $\pT$ values, 
parton scatterings receive major contributions from the region of low fractional momenta 
($x=p_{\mbox{\tiny{\it parton}}}/p_{\mbox{\tiny{\it hadron}}}$), 
where the gluon distribution rises very fast.
In this regime, around a ``saturation scale'' $Q_{sat}^2$ of a few GeV$^2$,
parton branching and fusion processes should start to compensate each other
saturating the growth of the PDFs as $x\to$~0~\cite{Gribov:1984tu}.\\

The MC event generators of high-energy hadronic collisions used in CR physics
-- such as \qgsjet 01 and II~\cite{Kalmykov:1997te,Ostapchenko:2004ss},
\sibyll~\cite{Engel:1992vf,Fletcher:1994bd,Ahn:2009wx} and \epos~\cite{Werner:2005jf} --
have evolved starting up from the RFT approach, based on Pomeron degrees of freedom
and thus naturally accounting for soft dynamics, generalized to include 
perturbative parton-parton processes via ``cut (hard) Pomerons'' diagrams. 
Multi-scattering phenomena (gluon saturation, MPI) are also implemented
through various procedures~\cite{DdE:2011}.

\section{LHC data versus cosmic ray MCs}
\label{sec:dataMC}

Extensive air showers initiated by interactions of primary cosmic-ray particles 
with air nuclei in the atmosphere constitute multi-step cascade processes 
involving electromagnetic and hadronic processes. The electromagnetic part 
is well described theoretically, whereas the hadronic interactions are 
modeled as summarized in the previous Section.
The most important EAS observables are the 
the depth in the atmosphere where the number of charged particles reaches its maximum 
$X_{\max}$(g/cm$^2$), the number of particles at maximum $N_{\max}$, 
and the number of electromagnetic particles ($e^\pm$, $\gamma$) 
and muons ($\mu^{\pm}$) at ground~\cite{Haungs:2003jv}.\\

The relation between high-energy hadronic interactions and EAS observables 
has been studied numerically in detail in~\cite{Ulrich:2010rg}. 
The depth of shower maximum $X_{\max}$ depends mainly on 
(i) the {\it inelastic cross section} ($\sigma_{inel}$) of the primary particle 
with air nuclei, (ii) the corresponding energy fraction ({\it inelasticity})
transferred to secondary particles but the most energetic ``leading''
one emitted at very forward rapidities, relative to the primary particle,
and (iii) the {\it multiplicity} ($N_{ch}$) of the primary and subsequent very high-energy interactions, 
which defines how the energy is distributed to secondary particles and corresponding 
sub-showers (and results in a given $N_{\max}$).  The rest of EAS properties at ground 
are closely related to $X_{\max}$ and $N_{\max}$.
We discuss below how the LHC measurements constrain the collision-energy evolution of 
quantities such as $\sigma_{inel}$, $N_{ch}$, or the inelasticity.

\subsection{Inelastic \pp\ cross section}

\begin{figure}[htbp]
\centering
\includegraphics[width=6.95cm]{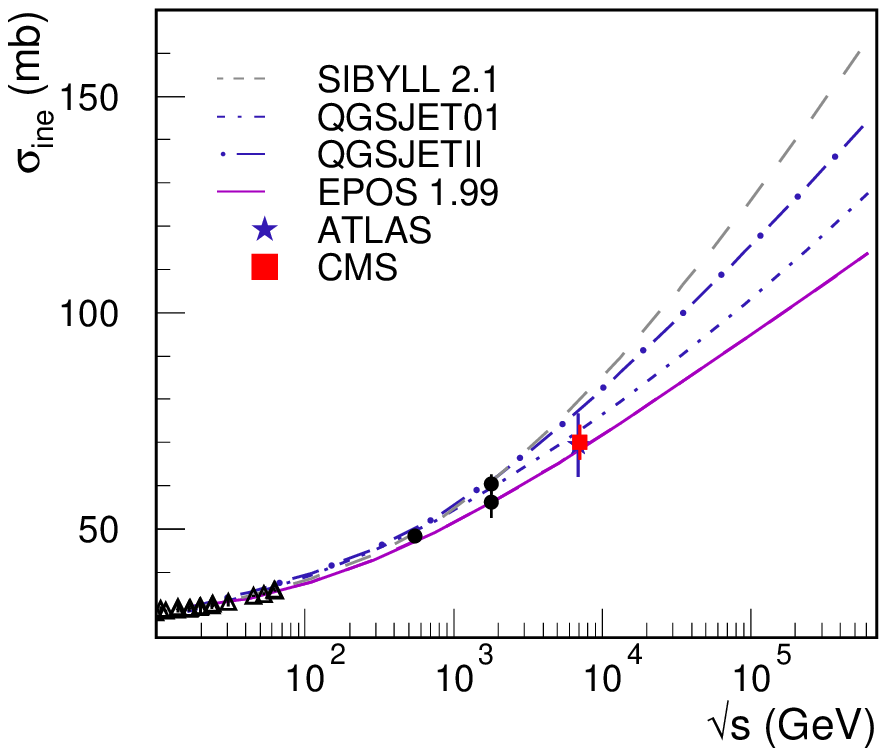}
\includegraphics[width=6.52cm]{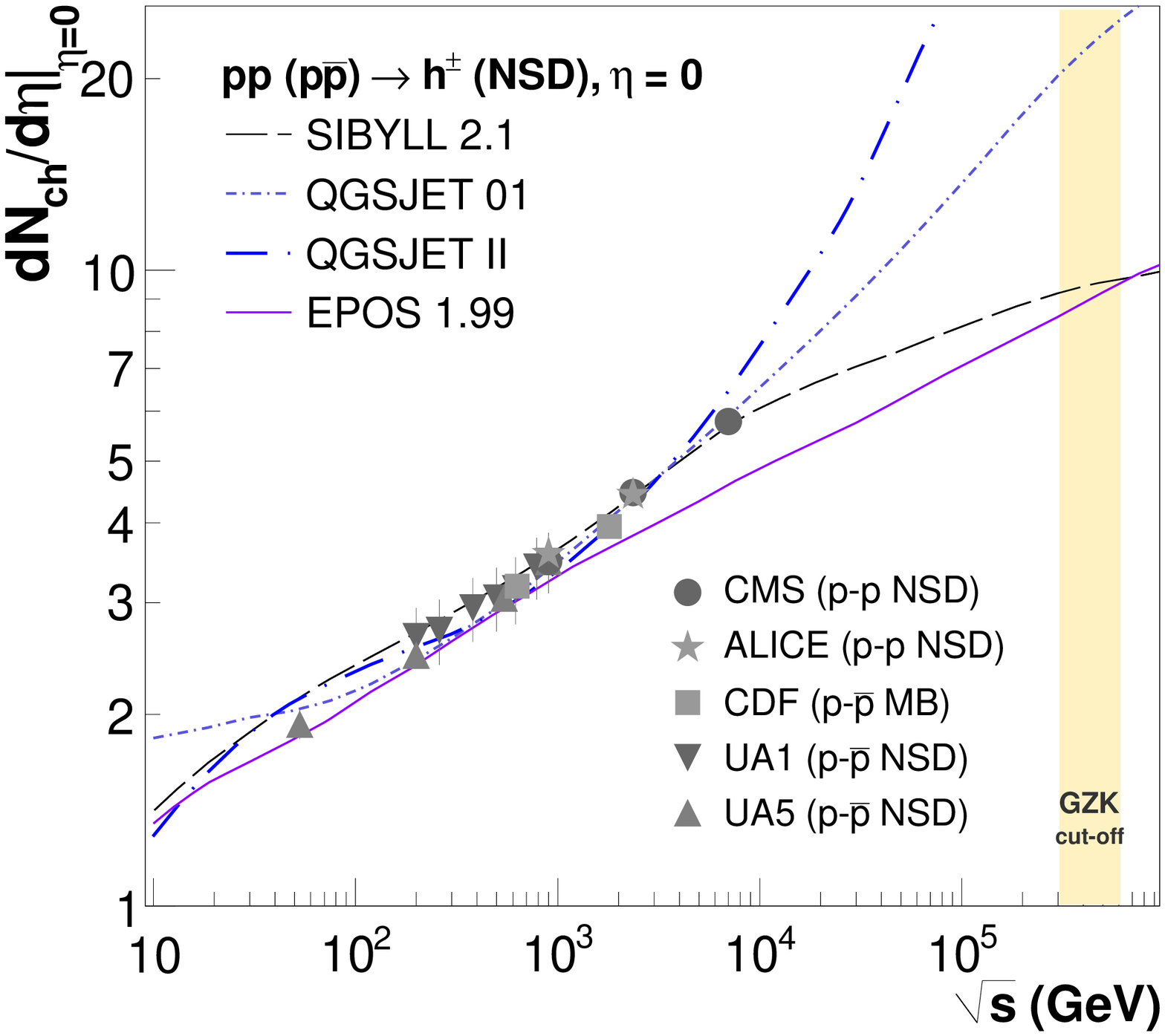}
\caption{Collision-energy dependence of the inelastic \pp(\ppbar) cross section
(left) and of the midrapidity charged hadron multiplicity density (right)
including the latest LHC data and cosmic-ray MCs predictions.}
\label{fig:sigmainel_dNdeta_vs_sqrts}
\end{figure}

A fundamental quantity of all CR models is the total hadronic cross section 
$\sigma_{tot}$ and its separation into elastic and inelastic (and, in particular,
diffractive) components. The measurement of $\sigma_{el}$ is accessible 
thanks to various forward proton detectors such as TOTEM~\cite{Anelli:2008zza} and 
ALFA (ATLAS)~\cite{Ask:2007fr} in the LHC tunnel area. The ATLAS~\cite{Aad:2011eu} 
and CMS~\cite{MaroneDIS11} experiments have already reported a value $\sigma_{inel}\approx$~70~mb
at $\sqrts$~=~7~TeV which seems to favour the lowest of the two (inconsistent) 
values previously measured at Tevatron (1.8~TeV) pointing to a slightly slower 
increase of the hadron production cross sections with c.m. energy as in \epos\ and \qgsjet 01
(Fig.~\ref{fig:sigmainel_dNdeta_vs_sqrts} left).

\subsection{Charged particle multiplicity}

The charged hadron pseudorapidity density at LHC energies provides an important constraint
for the modeling of the redistribution of energy in the first CR interactions in the atmosphere.
The midrapidity measurements of $\dNdeta\approx$~3.5,~4.5,~6 at $\sqrts$~=~0.9,~2.36,~7~TeV 
by ALICE~\cite{Aamodt:2010ft}, ATLAS~\cite{Aad:2010ir} and CMS~\cite{Khachatryan:2010xs}
indicate that the multiplicity changes smoothly in the lab energy range from $4\times 10^{14}$ 
to $3\times 10^{16}$\,eV, 
being well reproduced (with the exception of \epos) by the 
current interaction models used for EAS simulations (Fig.~\ref{fig:sigmainel_dNdeta_vs_sqrts}, right).
In addition, first results on the heavy-ion (Pb-Pb) multiplicity and its centrality dependence 
at $\sqrtsnn$~=~2.76~TeV~\cite{Collaboration:2010cz} provide extra important cross-checks 
on the role of initial-state gluon saturation effects in collisions involving nuclei, 
such as those of CR with air.

\subsection{Multiplicity probability distributions}

\begin{figure}[htbp]
\centering
\includegraphics[width=6.8cm,height=7.4cm]{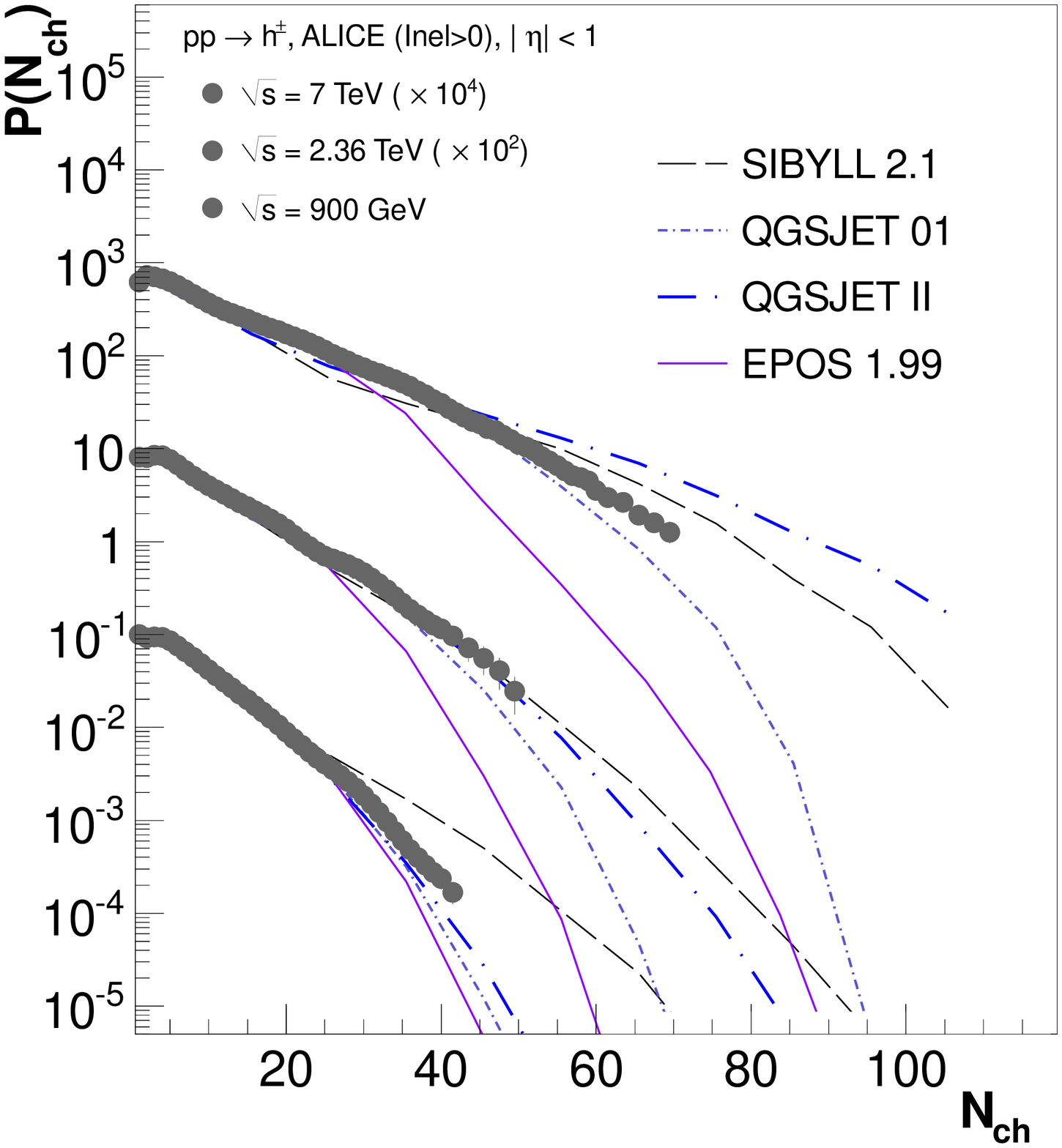}
\includegraphics[width=6.3cm,height=7.3cm]{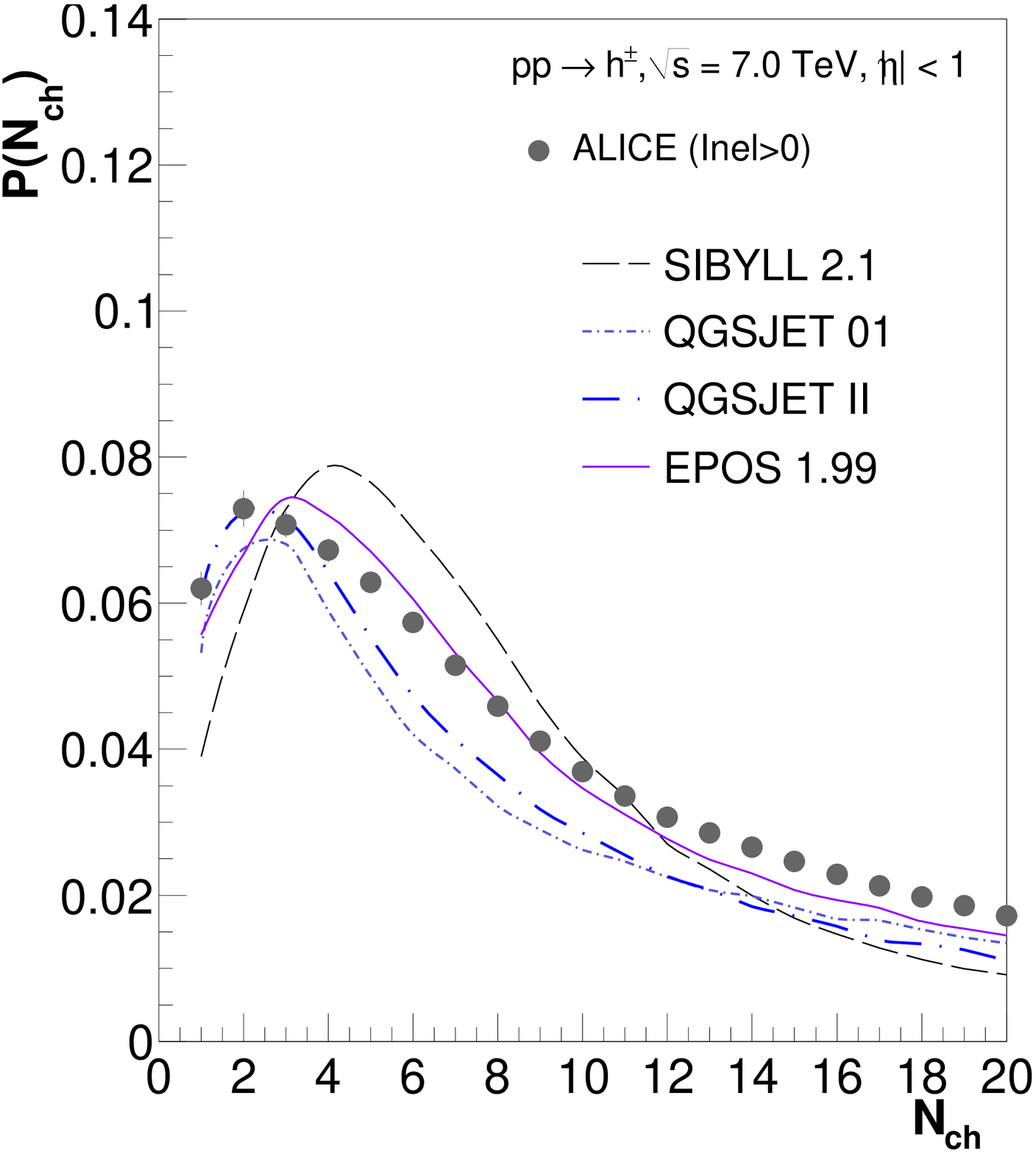}
\caption{Multiplicity distributions of charged hadrons, $P(N_{ch})$, measured by ALICE
in \pp\ events at $\sqrts$~=~0.9, 2.36 and 7 TeV~\cite{Aamodt:2010ft} (left) 
compared to \qgsjet 01 and II, \sibyll, and \epos.
The right plot shows a ``zoom'' in the low multiplicity range for $\sqrts$~=~7~TeV.}
\label{fig:mult_inel_vs_MC}
\end{figure}

The multiplicity distribution $P(N_{ch})$, i.e.~the probability to produce $N_{ch}$ charged
hadrons in an event provides important differential constraints on the internal details of 
the hadronic interaction models. The low multiplicity part 
is mostly dominated by the contributions from 
diffraction (single-cut Pomeron exchanges),
whereas the tail of the distribution gives information 
on the relative contribution of multiparton scatterings 
(multi-Pomeron exchanges). 
The experimental measurements (Fig.~\ref{fig:mult_inel_vs_MC})
at the three c.m.~energies measured so far at the LHC, indicate
that the high-$N_{ch}$ tail (left) is underestimated by \epos\ and \qgsjet01, 
whereas \sibyll\ and \qgsjet II get a bit closer, sometimes overestimating the data. 
In the low-$N_{ch}$ region around $P(N_{ch})\sim$~4 (right panel), only \epos\ globally 
reproduces the experimental results whereas the rest of the models overestimate the measurements 
up to +30\% for \sibyll. The peak is even shifted towards lower multiplicity in the case of 
both \qgsjet\ models. Thus, even if the average \pp\ multiplicities 
are well reproduced by most models (Fig.~\ref{fig:sigmainel_dNdeta_vs_sqrts} right), 
the details of their probability distributions are missed and indicate possible paths 
for improvement of the different model ingredients.

\subsection{Forward energy flow}

\begin{figure}[htbp]
\centering
\includegraphics[width=6.8cm,height=5.9cm]{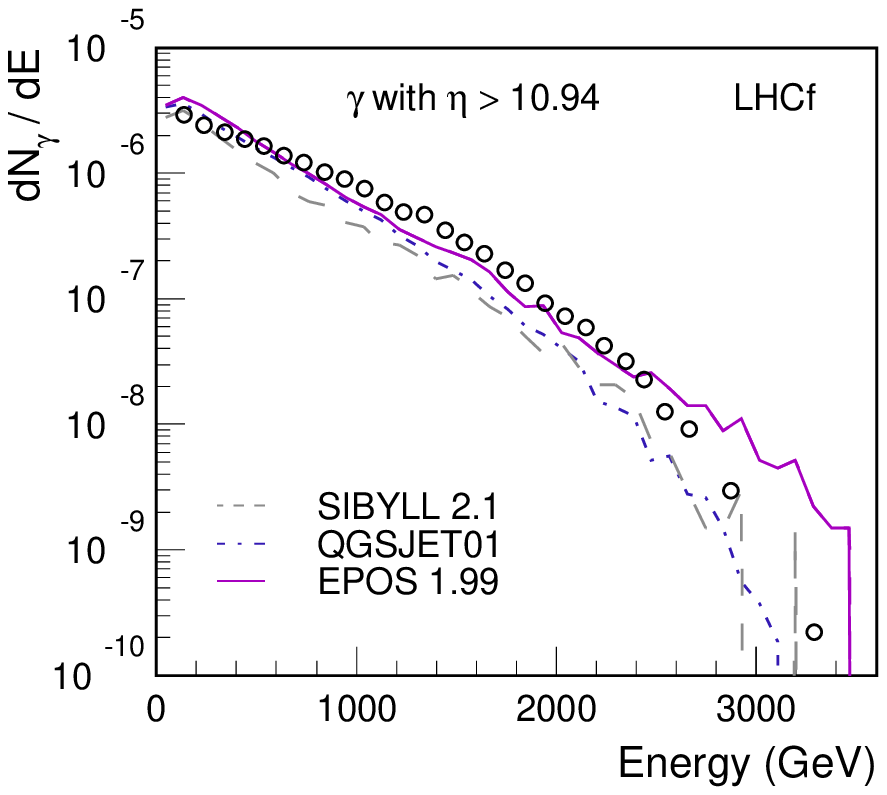}\hspace{0.4cm}
\includegraphics[width=6.4cm,height=6.cm]{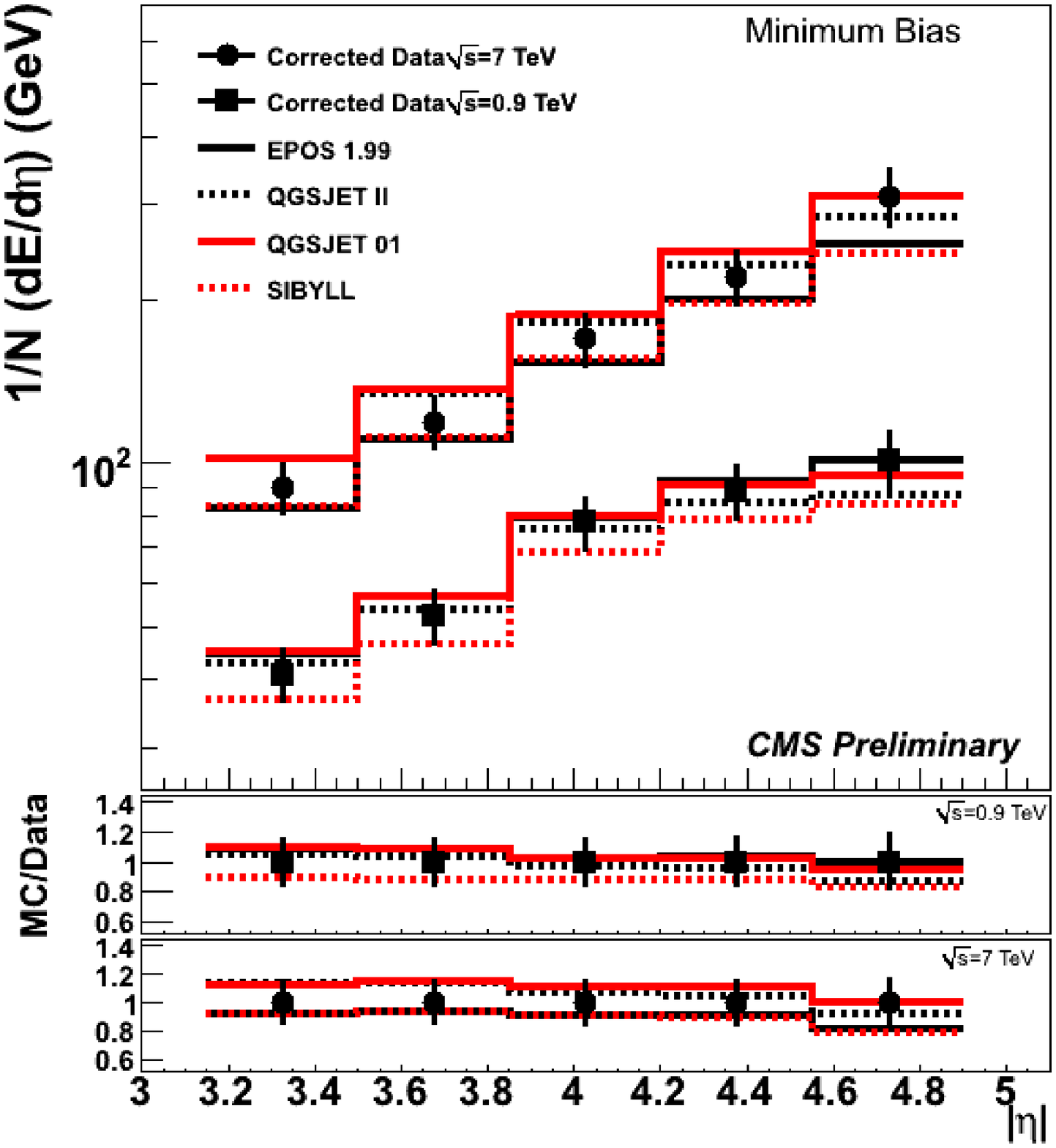}
\caption{Forward energy distributions in \pp\ collisions at the LHC compared to CR Monte Carlos,
for photons at $|\eta|>10.94$ measured by LHCf at 7~TeV~\cite{Adriani:2011} (left) 
and for inclusive hadrons at $3<|\eta|<5$ measured by CMS at 0.9, 7~TeV~\cite{KnutssonDIS11}
(right).}
\label{fig:fwd_energy}
\end{figure}

A key quantity for air shower development is the distribution of neutral energy 
emitted at very forward rapidities as it provides constraints on the production of 
leading hadrons (inelasticity) as well as on the transfer of energy from 
the hadronic core to the electromagnetic cascade (via $\pi^0\to\gamma\gamma$ decays).
The recent LHCf measurement of the $\gamma$ spectrum for rapidities above 
$|\eta|\approx 8.8$~\cite{Adriani:2011} is compared to model simulations
in Fig.~\ref{fig:fwd_energy} (left)~\cite{Pierog2011}.
The simulations are in relatively good agreement with the data within the systematical
uncertainty (not shown here) although for $E_\gamma\lesssim$~1.5~TeV the spectrum 
slope is harder in the data than in the predictions.
Of similar importance is the measurement of the energy flow and particle 
spectra in the forward range  $|\eta|$~=~5~--~10~\cite{dEnterria:2008jk}. 
This is an angular range that has been historically very difficult to access in 
collider experiments but that is partially covered by various detectors at the LHC
such as TOTEM~\cite{Anelli:2008zza} and CASTOR (CMS)~\cite{Andreev:2010zzb}.
Preliminary results of the energy flow in the $|\eta|$~=~3~--~5 range
indicate a good data--model agreement at $\sqrts$~=~0.9 and 7~TeV~\cite{KnutssonDIS11} (Fig.~\ref{fig:fwd_energy} right).


\subsection{Transverse momenta of hadrons}

Although the transverse momentum spectrum of the produced hadrons (or its average $\pT$)
at the LHC does not have a direct impact on the interpretation of air shower data
-- the lateral distributions of particles at ground is rather defined by multiple Coulomb 
scattering and by the $\pT$ spectra of secondaries at much lower energies -- 
such a measurement is of importance for checking the overall physics consistency of soft 
and hard interaction mechanisms implemented in the models. Indeed, at high energies 
the peak of the perturbative production comes from interactions between partons whose 
transverse momentum is around the saturation scale, $\pT \sim Q_{sat}$, producing (mini)jets 
of a few GeV which fragment into hadrons. In models with saturation of parton densities, 
the mean transverse momentum of the produced hadrons is of the order of the saturation 
scale $Q_{sat}$ in the high-energy limit.\\

\begin{figure}[htbp]
\centering
\includegraphics[width=6.35cm]{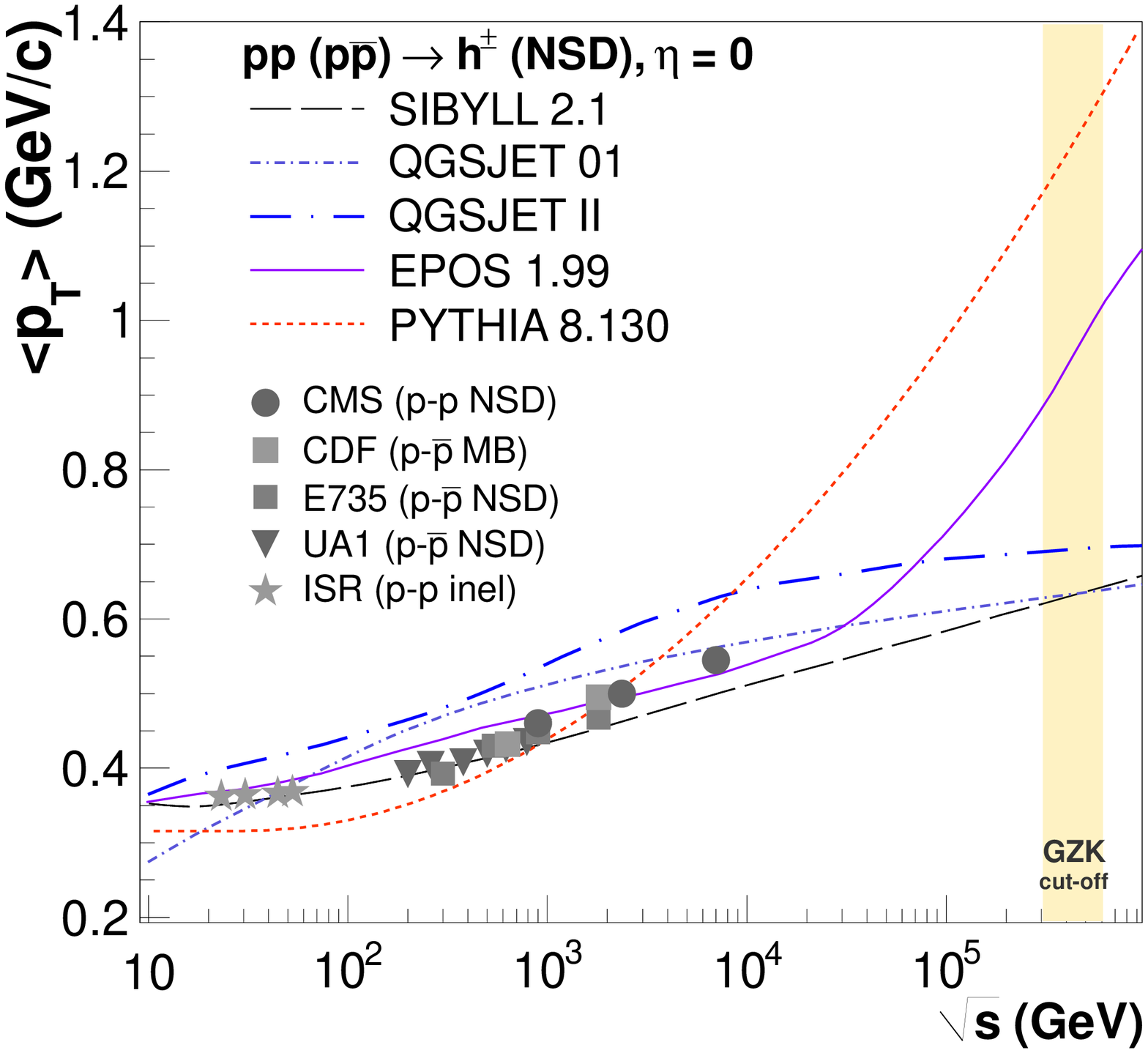}\hspace{0.2cm}
\includegraphics[width=6.99cm]{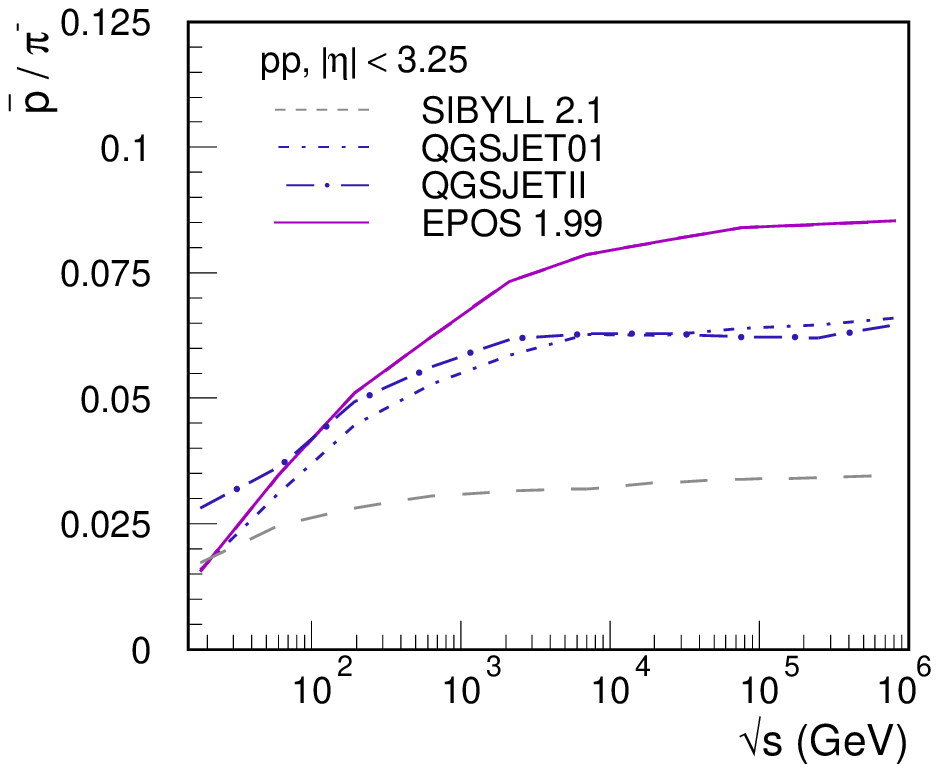}
\caption{Left: Average $\pT$ of charged particles at midrapidity in \pp\ (\ppbar) collisions 
as a function of $\sqrts$ compared to the predictions of CR event generators and \pythia\ 8.
Right: Collision-energy dependence of the ratio of antiprotons to $\pi^-$ yields at midrapidity 
predicted by various CR models.}
\label{fig:meanpT_baryon}
\end{figure}

In Fig.~\ref{fig:meanpT_baryon} (left) we show the energy evolution of the mean 
$\pT$ measured experimentally compared to the CR event generators and to 
the \pythia\ 8 MC~\cite{Sjostrand:2007gs}. 
All the RFT MCs but \epos\ predict a very moderate increase of $\meanpt$ with energy, 
reaching $\meanpt\sim$~0.6~GeV/c at GZK energies which is only 0.05~GeV/c above the current 
CMS result at 7 TeV, reflecting the moderate assumptions made on the saturation of the 
low-$x$ parton densities. \epos\ predicts a significantly larger $\meanptgzk\approx$~1~GeV/c
due to the inclusion of 
final-state collective parton expansion effects.
\pythia\ 8 -- whose dynamics is dominated by (mini)jet production with a running
$\pT$ cutoff that mimics parton saturation effects~\cite{DdE:2011} -- predicts a
higher average $\meanptgzk\approx$~1.2~GeV/c.\\

The level of (dis)agreement between the inclusive hadron observables measured at 
the LHC discussed in~\cite{DdE:2011,Pierog2011} and each one of the four CR hadronic 
MCs considered, is summarized in Table~\ref{tab:rft_summary}. 
The event generators give an overall decent description of all measurements but 
model improvements, particularly, those related to the treatment of 
inelastic diffraction and of the parton saturation mechanism, are desirable.

\begin{table}[htbp]
\begin{center}
\begin{tabular}{l|c@{}@{}c@{}@{}c|c@{}@{}c@{}@{}c|c@{}@{}c@{}@{}c|c@{}@{}c@{}@{}c@{}@{}c}\hline
\hspace{2.2cm}\footnotesize{Model} & & \sibyll\ 2.1 &&& \qgsjet 01 &&& \qgsjet II &&& \epos\ 1.99 & \\ 
\hspace{1.9cm}$\sqrts$ (TeV) & 0.9 & 2.36 & 7 & 0.9 & 2.36 & 7 & 0.9 & 2.36 & 7 & 0.9 & 2.36 & 7 \\ \hline
$\sigma_{inel}$ & \checkmark&$\Uparrow$&$\Uparrow$& \checkmark&\checkmark&\checkmark & \checkmark&$\Uparrow$&$\Uparrow$ & \checkmark&\checkmark&\checkmark \\
$\dNdeta$ & \checkmark&\checkmark&\checkmark & \checkmark&\checkmark&\checkmark & \checkmark&\checkmark&$\Uparrow$ & \checkmark&$\Downarrow$&$\Downarrow$ \\
$P(N_{ch}< 5)$  & $\Uparrow$&$\Uparrow$&$\Uparrow$ & $\Uparrow$&$\Uparrow$&$\Downarrow$ & $\Uparrow$&$\Uparrow$&$\Uparrow$ & \checkmark&\checkmark&\checkmark \\
$P(N_{ch}>30)$ & $\Uparrow$&\checkmark&$\Uparrow$ & \checkmark&$\Downarrow$&$\Downarrow$ & \checkmark&\checkmark&$\Uparrow$ & $\Downarrow$&$\Downarrow$&$\Downarrow$ \\
$\meanpt$ & \checkmark&$\Downarrow$&$\Downarrow$ & $\Uparrow$&$\Uparrow$&\checkmark & $\Uparrow$&$\Uparrow$&$\Uparrow$ & \checkmark&\checkmark&\checkmark \\
 \hline
\end{tabular}
\caption{Level of overall agreement between \qgsjet 01, \qgsjet II, \sibyll\ 2.1 and \epos\ 1.99
with inclusive charged hadron results measured in collisions 
at 0.9, 2.36 and 7~TeV: inelastic cross section $\sigma_{inel}$, pseudorapidity densities $\dNdeta$, 
multiplicity probabilities $P(N_{ch})$ for low and high values of $N_{ch}$, 
and mean transverse momentum $\meanpt$. A tick (\checkmark) indicates
a reasonable data--model agreement within experimental uncertainties, and $\Uparrow$  ($\Downarrow$) 
that the MC tends to over (under) estimate the measurements.} 
\label{tab:rft_summary}
\end{center}
\end{table}

\section{Conclusions}

Event generators used in high-energy cosmic ray (CR) physics include a 
description of hadronic interactions that partially
depends on non- and semi-perturbative QCD dynamics that need to be 
calibrated with experimental data. The highest energy cosmic rays measured 
on Earth at $E_{lab}\approx$~10$^{20}$~GeV collide with the nuclei in the 
atmosphere at c.m. energies more than two orders of magnitude above those 
studied at particle colliders before the LHC. The most recent results from \pp\ 
collisions at the LHC (equivalent to lab energies around $E_{lab} = 3\times 10^{16}$\,eV) 
are of big help to constrain the details of multiparticle production in the 
Monte Carlos used to describe CR air showers. 
The measured characteristics of the bulk of hadron production at multi-TeV 
energies does not reveal serious deficiencies in any of the models.
This gives a strong support to the interpretation of the results 
in the CR ``knee'' energy range ($E_{lab} = 10^{15.5}$eV) in terms of conventional
primary spectrum and nuclear mass composition and disfavours some proposed
speculative ideas that the change of the CR spectral slope 
could be due to a sudden change in the hadronic interaction mechanism 
above $2$\,TeV c.m.\ energy (see e.g.~\cite{Barcelo:2009uy}).\\

Although the first LHC measurements support a conventional extrapolation 
of the known features of multiparticle production to the highest known energies,
none of the models is in perfect agreement with all the hadronic observables
measured at the LHC (see Table~\ref{tab:rft_summary}). In particular, 
extrapolations at the GZK-cutoff energies span a range of predictions -- 
e.g. $\dNdeta\approx$~10 (\epos, \sibyll)~--~50~(\qgsjet II) for the particle densities
and $\meanpt\approx$~0.6~(\sibyll, \qgsjet 01)~--~1~(\epos)~GeV/c for the 
mean hadron transverse momentum -- that
justify the concurrent use of various MCs to gauge the uncertainties connected 
to hadronic interaction models in the interpretation of the cosmic ray
data. 
New retuning of model parameters and reconsideration of model assumptions, e.g.
for \epos~\cite{Pierog:2010dt} and \qgsjet~\cite{Ostapchenko:2010vb},
are currently underway.\\

Further improvement of our understanding of the strong interaction and of 
the properties of cosmic-rays at multi-TeV energies 
will be provided by the LHC with \pp\ data at the nominal c.m.~energy of 
$\sqrts$~=~14\,TeV (corresponding to CR protons of $10^{17}$\,eV in the lab frame),
as well as from
the expected proton-nucleus ($p$-$Pb$) runs at $\sqrtsnn$~=~8.8~TeV~\cite{Salgado:2011wc},
since the CR-induced air showers in the upper atmosphere are mostly from (p,$\alpha$,Fe)+(N,O) collisions. 
Also, since baryon-induced subshowers lead to a higher number of muons at ground than 
meson-induced ones, the energy dependence of the baryon production rate (see e.g. Fig.~\ref{fig:meanpT_baryon}
right), its relation to the centrality of the collision, and the momentum distribution
of the baryons constitute also important quantities to be measured.



\begin{thebibliography}{}
%
%


\bibitem{Greisen:1966jv}
K.~Greisen,
Phys. Rev. Lett. 16 (1966) 748--750.

\bibitem{Zatsepin66e}
G.~T. Zatsepin and V.~A. Kuzmin,
J. Exp. Theor. Phys. Lett. 4 (1966) 78.

\bibitem{Bluemer:2009zf}
J.~Bl{\"u}mer, R.~Engel, and J.~R. H{\"o}randel,
Prog. Part. Nucl. Phys. 63 (2009) 293--338

\bibitem{Knapp:2002vs}
J.~Knapp, D.~Heck, S.~J. Sciutto, M.~T. Dova, and M.~Risse,
Astropart. Phys. 19 (2003) 77--99

\bibitem{DdE:2011}
D. d'Enterria, R. Engel, T. Pierog, S. Ostapchenko and K. Werner, 
Astropart. Phys. to appear; arXiv:1101.5596

\bibitem{Pierog2011}
T. Pierog, D. d'Enterria, R. Engel, S. Ostapchenko and K. Werner, 
Proceeds. ICRC'11.

\bibitem{Gribov:1968fc}
V.~N. Gribov,
Sov. Phys. JETP 26 (1968) 414--422.

\bibitem{Gribov:1972ri}
V.~N. Gribov and L.~N. Lipatov,
Sov. J. Nucl. Phys. 15 (1972) 438--450.

\bibitem{Altarelli77}
G.~Altarelli and G.~Parisi,
 Nucl.~Phys.~B126 (1977) 298.

\bibitem{Dokshitzer77}
Yu.~L.~Dokshitzer,
 Sov.~Phys.~JETP~46 (1977) 641.

\bibitem{Andersson:1983ia}
B.~Andersson, G.~Gustafson, G.~Ingelman, and T.~Sjostrand,
Phys. Rept. 97 (1983) 31.

\bibitem{Gribov:1984tu}
L.~V. Gribov, E.~M. Levin, and M.~G. Ryskin,
Phys. Rept. 100 (1983) 1--150.

\bibitem{Kalmykov:1997te}
N.~N. Kalmykov, S.~S. Ostapchenko, and A.~I. Pavlov,
Nucl. Phys. Proc. Suppl. 52B (1997) 17--28.


\bibitem{Ostapchenko:2004ss}
 S.~Ostapchenko,
Nucl.~Phys.~Proc.~Suppl.~151 (2006) 143-146 
     

\bibitem{Engel:1992vf}
J.~Engel, T.~K. Gaisser, T.~Stanev, and P.~Lipari,
Phys. Rev. D46 (1992) 5013--5025.

\bibitem{Fletcher:1994bd}
R.~S. Fletcher, T.~K. Gaisser, P.~Lipari, and T.~Stanev,
Phys. Rev. D50 (1994) 5710--5731.

\bibitem{Ahn:2009wx}
E.-J. Ahn, R.~Engel, T.~K. Gaisser, P.~Lipari, and T.~Stanev,
Phys. Rev. D 80 (2009) 094003

\bibitem{Werner:2005jf}
K.~Werner, F.-M. Liu, and T.~Pierog,
Phys. Rev. C74 (2006) 044902

\bibitem{Haungs:2003jv}
A.~Haungs, H.~Rebel, and M.~Roth,
Rept. Prog. Phys. 66 (2003) 1145--1206.

\bibitem{Ulrich:2010rg}
R.~Ulrich, R.~Engel, and M.~Unger,
Phys. Rev. D 83 (2011) 054026 and arXiv:1010.4310 [hep-ph].

\bibitem{Anelli:2008zza}
G.~Anelli {\it et~al.}  (TOTEM Collab.),
JINST 3 (2008) S08007.

\bibitem{Ask:2007fr} S.~Ask,  arXiv:0706.0644 [hep-ex].  

\bibitem{Aad:2011eu}
  G.~Aad {\it et al.}  [ATLAS Collaboration],
  arXiv:1104.0326 [hep-ex].

\bibitem{MaroneDIS11} M. Marone (CMS Collab.), Proceeds. DIS'11

\bibitem{Aamodt:2010ft}
K.~Aamodt {\it et~al.}  (ALICE Collab.),
 Eur.\ Phys.\ J.\  C {\bf 68} (2010) 89; and
  Eur.\ Phys.\ J.\  C {\bf 68} (2010) 345

\bibitem{Aad:2010ir}
 G.~Aad {\it et~al.}  (ATLAS Collab.),
  arXiv:1012.5104 [hep-ex].

\bibitem{Khachatryan:2010xs}
V.~Khachatryan {\it et~al.}  (CMS Collab.),
JHEP 02 (2010) 041; and
Phys. Rev. Lett. 105 (2010) 022002

\bibitem{Collaboration:2010cz}
K.~Aamodt {\it et~al.}  (ALICE Collab.), Phys. Rev. Lett. 106 (2011) 032301 

\bibitem{Adriani:2011}
O.~Adriani {\it et~al.} (LHCf Collab.)
arXiv:1104.5294 [hep-ex].

\bibitem{dEnterria:2008jk}
D.~d'Enterria, R.~Engel, T.~McCauley, and T.~Pierog, Indian J. Phys. 84 (2010) 1837;
arXiv:0806.0944.

\bibitem{Andreev:2010zzb} V.~Andreev {\it et al.},  Eur.\ Phys.\ J.\  C {\bf 67} (2010) 601. 

\bibitem{KnutssonDIS11} A. Knutsson (CMS Collab.), Proceeds. DIS'11

\bibitem{Sjostrand:2007gs}
T.~Sjostrand, S.~Mrenna, and P.~Z. Skands,
Comput. Phys. Commun. 178 (2008) 852--867; arXiv:0710.3820 [hep-ph].


\bibitem{Barcelo:2009uy}
R.~Barcel\'o, M.~Masip, and I.~Mastromatteo, JCAP 0906 (2009) 027; 
arXiv:0903.5247 [hep-ph].

\bibitem{Pierog:2010dt}
T.~Pierog, I.~Karpenko, S.~Porteboeuf, and K.~Werner,
arXiv:1011.3748 [astro-ph.HE].

\bibitem{Ostapchenko:2010vb}
S.~Ostapchenko,
Phys.~Rev.~D83 (2011) 014018; arXiv:1010.1869 [hep-ph].

\bibitem{Salgado:2011wc}
  C.~A.~Salgado {\it et al.},
  arXiv:1105.3919 [hep-ph].

\end{thebibliography}


\end{document}